\documentclass[twoside,english,preprint,5p]{elsarticle}
\usepackage[T1]{fontenc}
\usepackage[latin9]{inputenc}
\usepackage{color}
\usepackage{babel}
\usepackage{booktabs}
\usepackage{amstext}
\usepackage{graphicx}
\usepackage{esint}
\usepackage[unicode=true,
 bookmarks=false,
 breaklinks=false,pdfborder={0 0 1},backref=section,colorlinks=false]
 {hyperref}

\makeatletter

\providecommand{\tabularnewline}{\\}

\usepackage{babel}
\journal{Theor. Appl. Mech. Lett.}

\addto\captionsenglish{}

\makeatother

\begin{document}
\begin{frontmatter}{}

\title{{\Large{}Numerical simulation of vortex-induced drag of elastic swimmer
models}}

%kai: I put the title I sent the editor of TAML

\author{T.~Engels\fnref{TUB,ENS}}

\author{D.~Kolomenskiy\fnref{JPN}}

\author{K. Schneider\corref{cor1}\fnref{AMU}}

%kai: I put me as corresponding author, as I am going to submit the paper.

\cortext[cor1]{Corresponding author: kai.schneider@univ-amu.fr}

%\ead{kai.schneider@univ-amu.fr}
%\ead{}

\author{J.~Sesterhenn\fnref{TUB}}

%\ead[url]{http://www.elsevier.com}

\address[TUB]{ISTA, Technische Universität Berlin, Berlin, Müller-Breslau-Strasse
15, 10623 Berlin, Germany }

\address[ENS]{LMD-CNRS, Ecole Normale Supérieure, 24 rue Lhomond, 75231 Paris
Cedex 05, France }

\address[JPN]{CEIST, Japan Agency for Marine-Earth Science and Technology (JAMSTEC).
3173-25 Showa-machi, Kanazawa-ku, Yokohama Kanagawa 236-0001, Japan. }

\address[AMU]{Institut de Mathématiques de Marseille, CNRS, Aix-Marseille Université,
39 rue F. Joliot-Curie, 13453 Marseille Cedex 13, France }
\begin{abstract}
We present numerical simulations of simplified models for swimming
organisms or robots, using chordwise flexible elastic plates. We focus
on the tip vortices originating from three-dimensional effects due
to the finite span of the plate. These effects play an important role
when predicting the swimmer's cruising velocity, since they contribute
significantly to the drag force. First we simulate swimmers with rectangular
plates of different aspect ratio and compare the results with a recent
experimental study. Then we consider plates with expanding and contracting
shapes. We find the cruising velocity of the contracting swimmer to
be higher than the rectangular one, which in turn is higher than the
expanding one. We provide some evidence that this result is due to
the tip vortices interacting differently with the swimmer. 
\end{abstract}
\begin{keyword}
swimming \sep fluid-structure interaction \sep thrust generation
\sep numerical simulation 
\end{keyword}
\end{frontmatter}{}

\section{Introduction}

Swimming organisms exploit bending waves to produce propulsive force,
an effect which has been extensively studied. Predicting their cruising
velocity, however, remains challenging, as the drag force has to be
taken into account. In this work, we numerically simulate simplified
``swimmers'', which consist of a chordwise flexible plate \textcolor{black}{with
an imposed} pitching motion at the leading edge, immersed in a viscous,
incompressible fluid. The solid is \textcolor{black}{fully} coupled
\textcolor{black}{to} the fluid, i.e., we deal with a fluid\textendash structure
interaction problem. The emphasis is placed on the longitudinal tip
vortices, which result from the finite span of the plate, and their
contribution to the drag force.

The usage of flexible foils for thrust generation as a simplified
model for swimming organisms is common in both experimental and numerical
contributions. Dewey et al. \cite{Dewey2013} for instance studied
flexible pitching panels experimentally. They found the efficiency,
i.e., the ratio of thrust to power coefficient, to be maximized if
the Strouhal number is in the range $0.25<St<0.35$ and the pitching
frequency is tuned to the structural resonant frequency of the foil.
The former result is supported by a variety of contributions 
\cite{Eloy2012,Triantafyllou2004,Triantafyllou1993}.
The connection between the driving frequency $f$ and the resonant
frequency $f_{0}$ is subject to some controversy in the community.
Kang et al. \cite{Kang2011} state that operating at or near a structural
resonant will enhance performance, a fact which is widely accepted.
However, different studies found the precise relation $f/f_{0}$ to
vary appreciably. For example, Ramananarivo et al. \cite{Ramananarivo2011}
state optimal performance around $f/f_{0}=0.7$. Two-dimensional data
\cite{Kolomenskiy2013} points in the same direction, although the
difference to the resonant is smaller. Yeh and Alexeev \cite{Yeh2014}
found two regimes which maximize cruising speed and efficiency at
$f/f_{0}\approx1.1$ and $1.6$, respectively. However, they normalized
by the resonant frequency in fluid, which can be derived analytically
\cite{Sader2007}. Contrarily to these findings, Vanella et al. 
\cite{Vanella2009}
provided evidence for peak efficiency in flexible insect wings around
$0.33$. The proposed argument is the usage of superharmonic resonances,
also stated in \cite{Ramananarivo2011}. Collectively, these findings
indeed suggest an important role of the resonant frequency, though
the exact relation remains not fully understood.

The total drag acting on these swimming organisms or robots can be
decomposed into the contributions of the friction drag and the vortex
induced drag. The former contribution has been relatively well explored.
Theoretical studies have considered the laminar boundary layer, which
is either compressed or stretched by the undulatory motion of the
swimmer \cite{Lighthill1971}. This effect is usually referred to
as the ``Lighthill boundary-layer thinning hypothesis''. More recently,
Ehrenstein et al. \cite{Ehrenstein2014} employed high-quality numerical
simulations using body-fitted meshes to quantify and verify this hypothesis.

The vortex induced drag, which may play a significant role, has only
recently gained attention of experimentalists. In the context of simplified
mechanical swimming robots, Raspa et al. \cite{Raspa2014} established
a basic model to explain the influence of the finite aspect ratio
by the formation of trailing longitudinal tip-vortices. The present
numerical study is inspired by these experiments, and should be seen
as complimentary approach, given the difficulty of experimentally
measuring the instantaneous flow field appropriately. In a first step,
using rectangular swimmers, we will reproduce some experimental results
and confirm the interpretation that the tip vortices play a major
role in the drag force of the swimmer. In a second step, we move on
and modify the swimmer's shape and find that a contracting form may
be advantageous in terms of terminal cruising speed. 
\begin{figure}
\begin{centering}
\includegraphics[width=0.7\columnwidth]{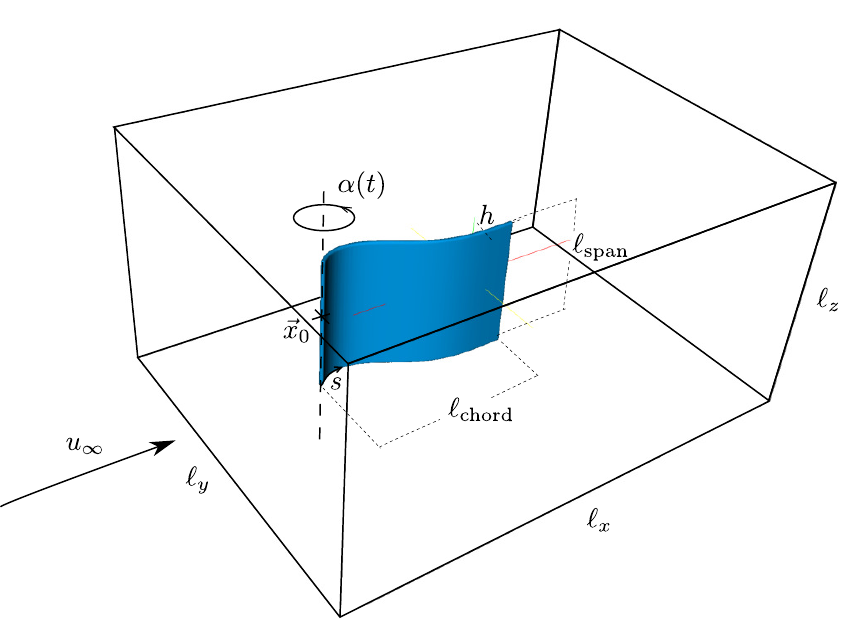} 
\par\end{centering}
\caption{\textit{Setup of the present work. The swimmer, which consists of
a %kai: this is wrong: spanwise flexible plate 
}\textit{\textcolor{black}{chordwise}}\textit{ flexible and spanwise
rigid plate undergoing an imposed pitching motion $\alpha\left(t\right)$,
is immersed in a viscous incompressible fluid with imposed or dynamically
computed axial mean flow $u_{\infty}$.\label{fig:Setup-used-in}}}
\end{figure}

\section{Materials and Methods}

The experimental swimmer used in \cite{Raspa2014} consists of a Mylar
sheet attached to a driving shaft, driven by a stepper motor. The
whole swimmer, including the motor, \textcolor{black}{can} move in
the $x$-direction. For the numerical simulations in the present work,
some assumptions \textcolor{black}{are }made in order to simplify the
problem's complexity sufficiently.

A\textcolor{black}{{} central} assumption of our numerical swimmer is
that the plate is perfectly rigid in the spanwise and flexible only
in the chordwise direction. Focusing on the chordwise flexibility
simplifies the complexity of the employed solid model; it is 1D. Experimental
findings show that spanwise deformations are present but rather small
in magnitude. \textcolor{black}{Here}, the swimmer is thus modeled
as a slender beam, made of linearly elastic, inextensible material,
\textcolor{black}{undergoing} large, non\textendash linear deformations. 

\textcolor{black}{We parametrize the beam with its local deflection
angle $\theta$ and its internal longitudinal force $T$. Integrating
the deflection line yields, for any value of $\theta$, automatically
the same length. In other words, we can take care of the inextensibility
condition by choosing this particular parametrization, since our basis
function satisfies it automatically. In those variables, the governing
equations read}
\begin{eqnarray}
\frac{\partial^{2}T}{\partial s^{2}}-T\left(\frac{\partial\theta}{\partial 
s}\right)^{2} & = & -2\eta\frac{\partial\theta}{\partial 
s}\frac{\partial^{3}\theta}{\partial 
s^{3}}-\mu\left(\dot{\theta}+\dot{\alpha}\right)^{2}\nonumber \\
 &  & -\eta\left(\frac{\partial^{2}\theta}{\partial 
s^{2}}\right)^{2}-\left[p\right]^{\pm}\frac{\partial\theta}{\partial 
s}\label{eq:BEQ1}\\
\mu\left(\ddot{\theta}+\ddot{\alpha}\right) & = & 
-\eta\frac{\partial^{4}\theta}{\partial s^{4}}+2\frac{\partial T}{\partial 
s}\frac{\partial\theta}{\partial s}-\frac{\partial\left[p\right]^{\pm}}{\partial 
s}\nonumber \\
 &  & +\left(T+\eta\left(\frac{\partial\theta}{\partial 
s}\right)^{2}\right)\frac{\partial^{2}\theta}{\partial s^{2}}\label{eq:BEQ2}
\end{eqnarray}
\textcolor{black}{together with the clamped-free boundary conditions}
\begin{eqnarray}
\left.\begin{array}{ccc}
\Theta & = & 0\\
\frac{\partial T}{\partial s}+\eta\frac{\partial^{2}\Theta}{\partial 
s^{2}}\frac{\partial\Theta}{\partial s} & = & 0\\
T\frac{\partial\Theta}{\partial s}-\eta\frac{\partial^{3}\Theta}{\partial s^{3}} 
& = & \left[p\right]^{\pm}
\end{array}\right\}  & \text{at clamped end}\label{eq:BC_BEQ}\\
\left.\begin{array}{ccc}
T & = & 0\\
\frac{\partial\Theta}{\partial s} & = & 0\\
\frac{\partial^{2}\Theta}{\partial s^{2}} & = & 0
\end{array}\right\}  & \text{at free end}\label{eq:BC_BEQ_2}
\end{eqnarray}
where $\alpha$ is the driven pitching motion, as illustrated in figure
\ref{fig:Setup-used-in}, $\left[p\right]^{\pm}$ is the pressure
jump across the beam, $s$ is the arclength coordinate, 
\textcolor{black}{$\mu=h\varrho_{s}/\ell\varrho_{f}$
is the dimensionless density and $\eta=EI/\ell^{3}\varrho_{f}U^{2}$
is the dimensionless stiffness.} The material properties reported
in \cite{Raspa2014} yield $\mu^{\mathrm{exp}}=0.0012$ and $\eta=0.0134$.
For numerical stability reasons, we set $\mu=0.0096$ instead, as
explained later. Two-dimensional simulations confirmed that the solution
is not very sensitive to the value of $\mu$ in this regime. The swimmer's
length $\ell_{\mathrm{chord}}=0.15\,\left[\mathrm{m}\right]$, the
fluid density $\varrho_{f}=1000\,\left[\mathrm{kg}/\mathrm{m}^{3}\right]$,
a time scale $T=1\,\left[\mathrm{s}\right]$ and the velocity scale
$U=\ell_{\mathrm{chord}}/T$ have been used for normalization. 
\textcolor{black}{Note
that we assume the beam clamped in a rotating relative system, which
rotates as $\alpha=\alpha_{\mathrm{max}}\sin\left(2\pi ft\right)$
with $\alpha_{\mathrm{max}}=50^{\circ}$. Therefore $\theta=0$ at
the leading edge. }Contrary to the experiment, we do not vary the
driving frequency $f$ but keep it fixed at unity (thus 
$f=1\,\left[\mathrm{Hz}\right]$).
\textcolor{black}{The solid model is similar to the ones used in 
\cite{Alben2009,Michelin2008}.
The non-linear terms stem from the geometric non-linearity.} The solid
model equations (\ref{eq:BEQ1}-\ref{eq:BC_BEQ_2}) are solved using
finite differences with an implicit time marching scheme, which treats
all terms, including the non-linear ones, implicitly. Details about
the solution procedure can be found in \cite{Engels2012a,Engels2014}.

The fluid is incompressible and Newtonian, and hence governed by the
Navier\textendash Stokes equations. To avoid using moving, body-fitted
meshes, the flexible structure is taken into account using the volume
penalization method. The governing penalized Navier\textendash Stokes
equations read 
\begin{eqnarray}
\partial_{t}\underline{u}+\underline{\omega}\times\underline{u} & = & -\nabla 
q+\frac{1}{\mathrm{Re}}\nabla^{2}\underline{u}-\frac{\chi}{C_{\eta}}
\left(\underline{u}-\underline{u}_{s}\right)\label{eq:NST_penal_primitive}\\
\nabla\cdot\underline{u} & = & 0\\
\underline{u}\left(\underline{x},t=0\right) & = & 
\underline{u}_{0}\left(\underline{x}\right).\label{eq:inicond2}
\end{eqnarray}
Note that eqns (\ref{eq:NST_penal_primitive}-\ref{eq:inicond2})
do not contain no-slip boundary conditions, since the geometric information
is encoded in the mask function $\chi(\underline{x},t)$ (where $\chi=0$
in the fluid and $\chi=1$ in the solid). \textcolor{black}{In the
present case, the mask thickness is set to four grid points, as problems
with vanishing thickness cannot be computed using this approach}.
The penalization parameter $C_{\eta}$ can be interpreted as solid
permeability and is chosen to a small value, here we use $C_{\eta}=10^{-3}$.
Details about the penalization method for flexible obstacles can be
found in \cite{Engels2012a,Engels2014}. \textcolor{black}{Guidelines
on how to choose $C_{\eta}$ are summarized in \cite{Engels2015a}.}
The mask function $\chi$ and the solid velocity field $\underline{u}_{s}$
are constructed from the solid model. 

The numerical solution of (\ref{eq:NST_penal_primitive}-\ref{eq:inicond2})
is obtained in a periodic domain using a Fourier pseudospectral method
and an explicit Adams\textendash Bashforth type time stepping 
\cite{Schneider2005}.
The grid is uniform and equidistant. \textcolor{black}{Solving the
Poisson equation required for the Helmholtz decomposition is done
in Fourier space, where the Laplace operator is diagonal and hence
the solution reduces to a simple division and no linear system has
to be solved. The in-house code for producing the results is open-source
and has been described in \cite{Engels2015a}; it can be run on supercomputers
using $\mathcal{O}\left(10\,000\right)$ CPUs}. 

Fluid-structure interaction problems are particularly challenging
since two non-linear PDEs are coupled. The most important parameter
for this coupling is the density ratio, which is defined by $\mu$
in the present article. The more similar the densities are, the more
challenging the simulation. Since this work deals with swimming, performed
in water, the simulations require an iterative coupling scheme. Still,
as the number of iterations depends significantly on $\mu$, we found
that the experimental value $\mu^{\mathrm{exp}}=0.0012$ requires
about 15-25 iterations (each at the price of one Navier\textendash Stokes
step), while the value we used, $\mu=0.0096$, requires only 3-5.
Two-dimensional simulations confirm that the difference in results
between both values is of a few per cent only, which justifies our
choice of $\mu$. 
\begin{figure}[ht!]
\begin{centering}
\includegraphics[width=0.8\columnwidth]{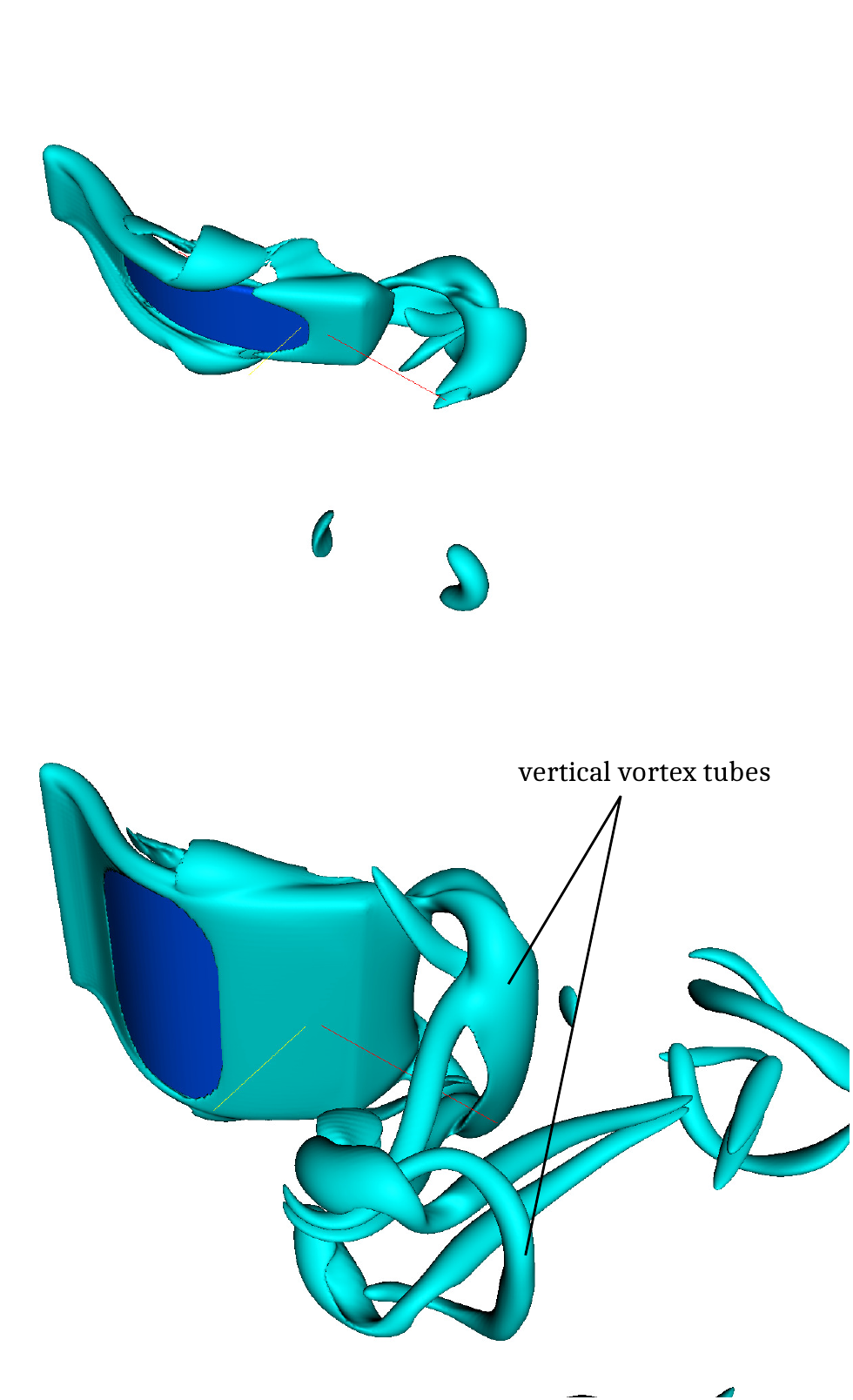} 
\par\end{centering}
\caption{\textit{Isosurfaces of vorticity $\left\Vert 
\underline{\omega}\right\Vert =17.5$
for aspect ratios $AR=0.2$ (top), $AR=0.7$ (bottom). For $AR=0.2$,
the tip vortices dominate the flow\label{fig:Isosurfaces-of-vorticity}}}
\end{figure}

\section{Rectangular Plates}

In the following we present the results obtained for rectangular plates.
The Reynolds number is $Re=U\ell_{\mathrm{chord}}/\nu=1000$, the
swimmer is computed in a box of size $2.66\times2.00\times1.33$ and
its leading edge at mid-span is located at 
$\underline{x}_{0}=(\begin{array}{ccc}
0.5, & 1.0, & 0.66\end{array})$. At this Reynolds number, we found a resolution 
of $512\times384\times256$
to be sufficient. The original experiment is performed at much higher
Reynolds number of $Re=22500$, which is currently out of scope for
numerical simulations. The value of the penalization parameter is
$C_{\eta}=10^{-3}$. The constant mean flow 
$\underline{u}_{\infty}=(\begin{array}{ccc}
0.5, & 0, & 0\end{array})$ is impulsively started at $t=0$, and we computed a 
total of $5$
periods. Since our discretization is periodic, a vorticity sponge
term is applied to all faces of the domain to prevent vortices from
re-entering the domain, with a parameter of $C_{\mathrm{sp}}=10^{-1}$,
see \cite{Engels2014} for more information about that boundary condition.
During the first period, the imposed pitching angle is multiplied
by a smooth startup conditioner, in order to avoid an impulsively
started motion, which would yield a pressure singularity. We carry
out 4 simulations with varying aspect ratio, 
$AR=\ell_{\mathrm{span}}/\ell_{\mathrm{chord}}=\{\begin{array}{cccc}
0.2, & 0.3, & 0.5, & 0.7\end{array}\}$. \textcolor{black}{In addition, a 
quasi-2D simulation has been performed,
in which the plate extends over the entire height of the domain. With
this simulation, it was verified that the }vortical structures are
stable at the $Re=1000$.

The vortical structure of the flow field is visualized in figure 
\ref{fig:Isosurfaces-of-vorticity}
for the smallest and largest values of $AR$, at the beginning of
the fourth stroke, $t=4.05$. In the $AR=0.7$ case, the vertical
vortex tubes can be observed. These tubes correspond to the vortices
shed in the 2D case. They connect to the tip vortices and form ring-like
structures, propagating perpendicular to the mean flow which also
advects them downstream. It is visible that, in the $AR=0.2$ case,
the tip vortices actually dominate the wake structure\textemdash the
vertical vortex tubes are not clearly distinguishable.

\begin{figure}
\begin{centering}
\includegraphics[width=0.65\columnwidth]{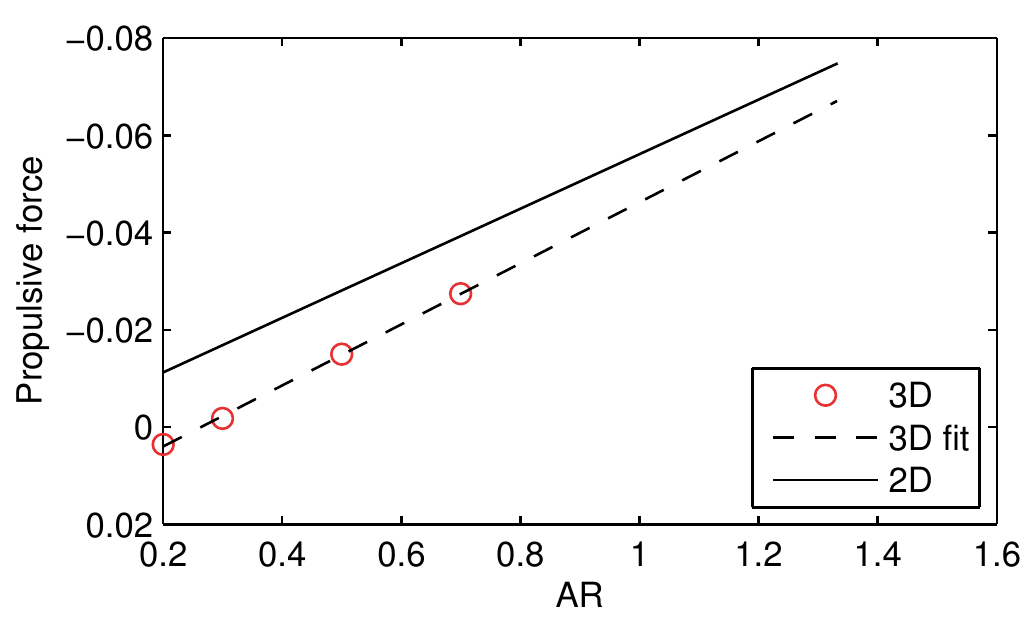} 
\par\end{centering}
\caption{\textit{Thrust force as a function of the aspect ratio. The solid
line represents the force predicted by the 2D approximation and the
dashed line is a linear least-squares fit through the available data
points from the 3D simulations.\label{fig:Thrust-force-as}}}
\end{figure}

The thrust force, that is the $x$-component of the hydrodynamic force,
is shown in figure \ref{fig:Thrust-force-as}. Note that thrust points
in negative $x$-direction. The solid line represents the prediction
based on 2D simulations. The four 3D simulations are marked by circles.
It can be observed that the thrust scales almost linearly with the
aspect ratio. \textcolor{black}{We} make the ansatz 
$F_{x}^{\mathrm{3D}}=F_{\mathrm{thrust}}\cdot AR+F_{\mathrm{tip}}$
and fit the coefficients $F_{\mathrm{thrust}}=-0.0628$ and 
$F_{\mathrm{tip}}=0.0165$
using least squares to the data points. \textcolor{black}{The 2D simulation
returns the force per unit span, $F_{\mathrm{thrust}}^{\mathrm{2D}}=-0.0561$,
and it excludes tip-vortices by definition. We can therefore estimate
the thrust as $F_{x}^{\mathrm{2D}}=F_{\mathrm{thrust}}^{\mathrm{2D}}\cdot AR$,
which is equivalent to a finite sized swimmer 
}\textcolor{black}{\emph{without}}\textcolor{black}{{}
tip-vortices.}

We can thus observe that the values for the thrust per unit span are
quite similar in both 3D and 2D cases, $F_{\mathrm{thrust}}\approx 
F_{\mathrm{thrust}}^{\mathrm{2D}}$,
and that the tip vortices indeed act like a constant offset. We can
thus numerically confirm the experimental results of Raspa et al.
\cite{Raspa2014}. 

\section{Non-rectangular Shapes}

The results for rectangular plates illustrate the importance of tip
vortices for the total drag. Actual fish on the other hand have of
course non-rectangular caudal fins, with possible consequences for
the vortical structures in the wake. We choose an additional set of
two different shapes, an expanding and a contracting form, to study
their influence on the cruising speed. All shapes have the same surface
and follow the same imposed driving motion. For simplicity, we still
assume the mechanical structure to be 1D and with constant $\mu$
and $\eta$, although the varying $\ell_{\mathrm{span}}\left(s\right)$
suggests that both should depend on $s$. This is a first order approximation,
since both $\mu,\eta$ are linear in $\ell_{\mathrm{span}}$, but
as $\eta\propto h^{3}$ the dominant effect of the stiffness $\eta$
is captured and as $\mu$ is small anyways (light swimmer), this assumption
seems justified. The non-rectangular shapes are defined as 
$\ell_{\mathrm{span}}^{\mathrm{exp}}\left(s\right)=2\left(\frac{0.35}{2}+0.525s^
{2}\right),\;\ell_{\mathrm{span}}^{\mathrm{contr}}\left(s\right)=2\left(\frac{
1.05}{2}-0.525s^{2}\right)$
and are illustrated  in figure \ref{fig:Different-shapes,-(a)}. 
\begin{figure}
\begin{centering}
\includegraphics[width=0.6\columnwidth]{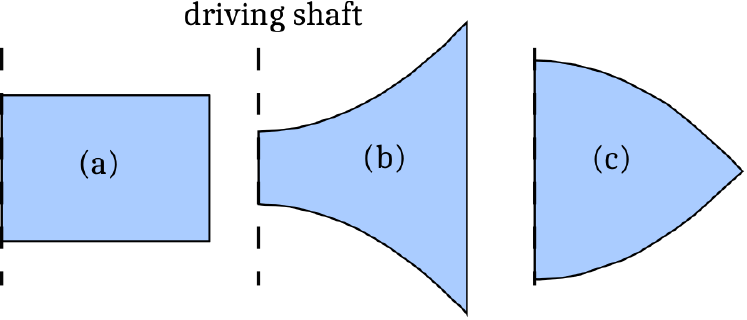} 
\par\end{centering}
\caption{\textit{Different shapes investigated here, termed (a) ``rectangular'',
(b) ``expanding'' and (c) ``contracting''. All three shapes have
the same surface\label{fig:Different-shapes,-(a)}.}}
\end{figure}

\begin{figure}
\begin{centering}
\includegraphics[width=1\columnwidth]{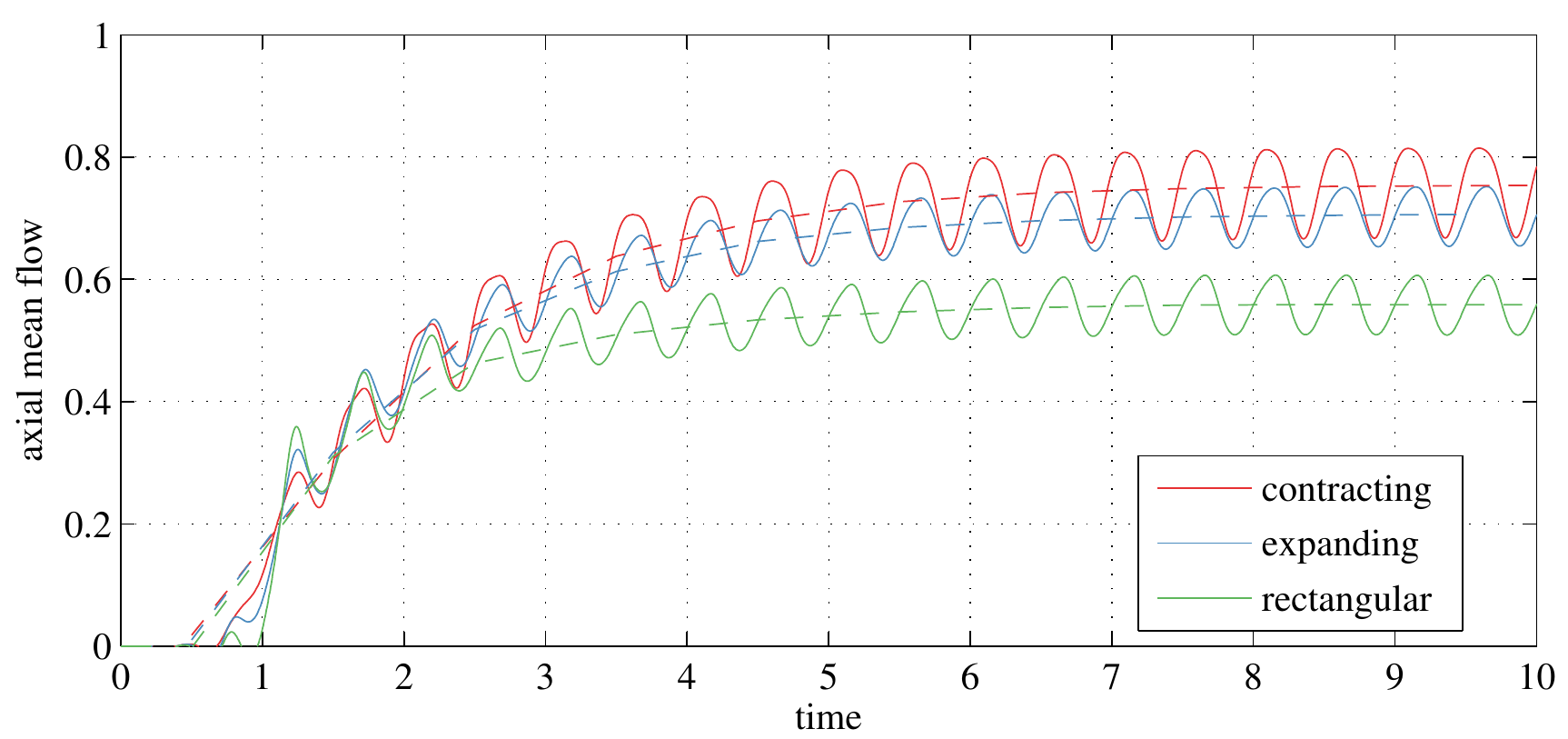} 
\par\end{centering}
\caption{\textit{Axial mean flow over time for the three different swimmers
from figure \ref{fig:Different-shapes,-(a)}. Solid lines is instantaneous
data, dashed lines is a moving average over the duration of one stroke.
\label{fig:Axial-mean-flow}}}
\end{figure}

\textcolor{black}{Our swimmers remain anchored to the laboratory system
and move the surrounding fluid instead. The mean flow in axial direction
is thus computed dynamically from $\dot{u}_{\infty}=\left\langle F\right\rangle 
/m_{\mathrm{fluid}}$,
where the initial condition is $u_{\infty}(t=0)=0$. In the steady
state, the mean flow balances the thrust force $\left\langle F\right\rangle $.}
The fluid mass is set to a relatively small value, $m_{\mathrm{fluid}}=0.1235$,
in order to speed-up the computation. \textcolor{black}{Note that the
cruising speed is not perfectly constant in the steady state, hence
the Galilean invariance is only satisfied approximately. }Figure 
\ref{fig:Axial-mean-flow}
illustrates the result obtained for all swimmers. They all reach their
steady state within 10 strokes, but the resulting cruising speed significantly
depends on the swimmer's shape. The contracting shape ($u_{\infty}=0.75$)
outruns both the rectangular ($u_{\infty}=0.70$) and the expanding
($u_{\infty}=0.55$) shapes. 

One remarkable difference between the three simulations is that the
expanding one has the smallest trailing edge displacement, which is
due to the larger concentration of area there. The pressure acting
on the tail is thus much higher in that case, reducing the deflection
amplitude. 

All swimmers have finite span and thus exhibit tip vortices, and again
these vortices offer a potential explanation for the higher cruising
velocity of the contracting shape. \textit{A priori}, one might think
the expanding form is advantageous, since the larger trailing edge
will produce larger vertical tube vortices (cf figure 
\ref{fig:Isosurfaces-of-vorticity}),
and thus reduce the spurious three-dimensional effects.

\begin{figure}[ht!]
\begin{centering}
\includegraphics[width=0.8\columnwidth]{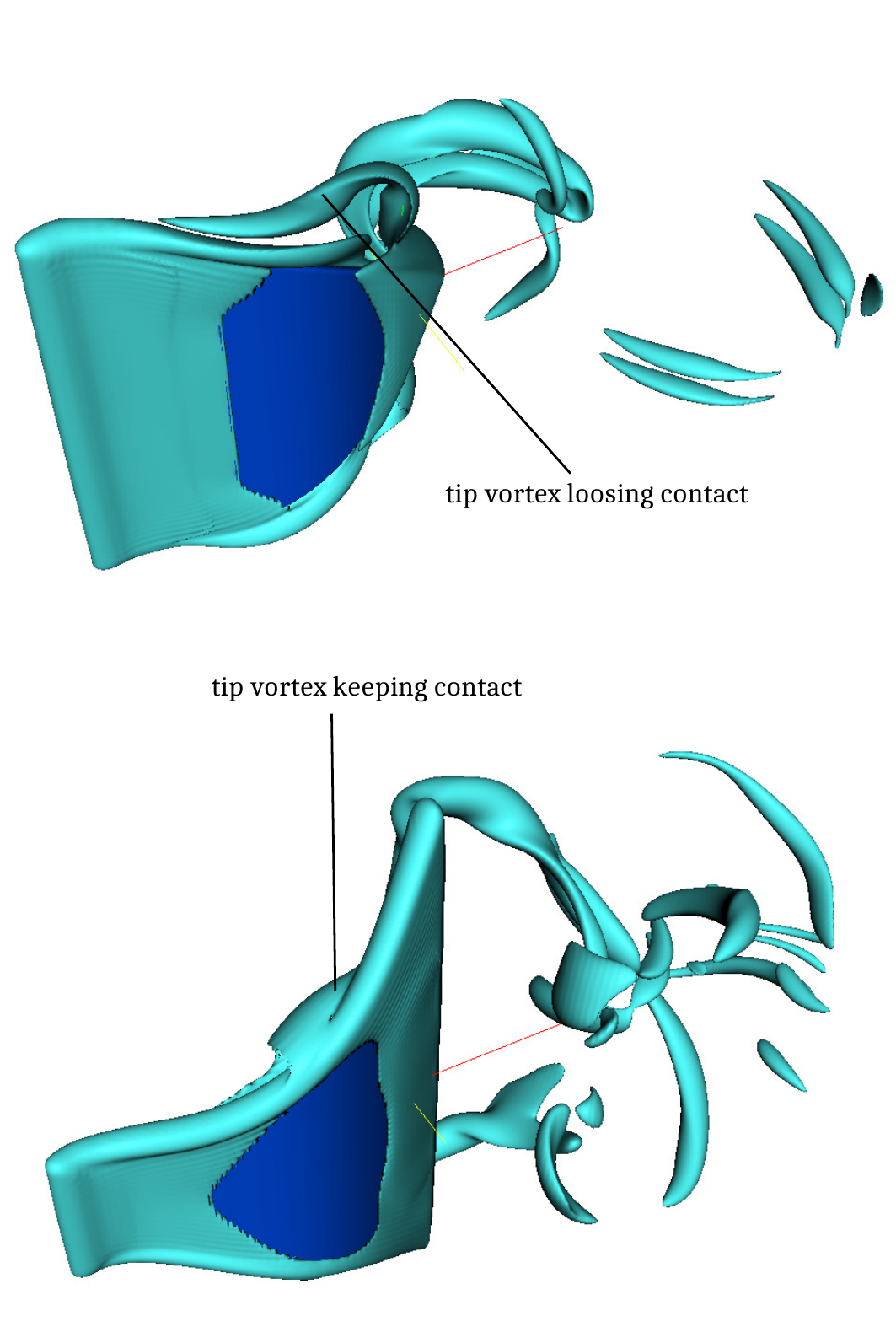} 
\par\end{centering}
\caption{\textit{Isosurfaces of vorticity $\left\Vert 
\underline{\omega}\right\Vert =17.5$
for contracting (top) and expanding shapes (bottom). }\label{fig:nice_fish}}
\end{figure}

However, the opposite is true. Figure \ref{fig:nice_fish} shows the
vortical structures for the contracting and expanding shape at the
same time, which is during the steady cruising state. The tip vortices,
shed in both configurations around mid-chord, are advected downstream
due to the mean flow, and they can be associated with a zone of lower
pressure. This drop in pressure creates a local net force pointing
in the direction of the vortex core, and part of which contributes
to the total drag force (depending on the orientation of the surface
normal relative to the $x$-direction). Visibly, in the contracting
case, the tip vortex quickly loses contact with the actual swimmer
\textendash{} its \textcolor{black}{tip-vortex induced} drag is thereby
reduced. The opposite is true for the expanding type swimmer: not
only does the tip vortex not lose contact with the swimmer, it does
instead even increase the portion of the swimmer influenced by the
tip vortices, compared to the rectangular swimmer. It can also be
noted that the total mean enstrophy, $\left\langle Z\right\rangle =\left\langle 
\iiint\left\Vert \underline{\omega}\right\Vert ^{2}d\underline{x}\right\rangle 
,$
which is a measure for the dissipation in the fluid wake, is significantly
higher in the expanding than in the contracting case, $\left\langle 
Z\right\rangle _{\mathrm{contracting}}=98.7$
versus $\left\langle Z\right\rangle _{\mathrm{expanding}}=127.7$,
indicating a higher dissipation rate in the expanding case. 

\section{Conclusion}

We numerically simulated the flow past simplified elastic swimmer
models. These models consist of flexible plates that have a driven
pitching angle at their leading edges. In the first part, we simulated
rectangular swimmers, that are inspired by the experimental work presented
in \cite{Raspa2014}. We \textcolor{black}{support} the finding that
the tip vortices, \textcolor{black}{or, more generally, 3D effects,
significantly contribute} to the total drag, and thus should be taken
into account when predicting the cruising speed of these swimmers.
\textcolor{black}{We illustrated that indeed the 3D nature of these
flows may act} like an \textcolor{black}{tip-vortex induced} drag that
is virtually independent of the aspect ratio. In a second step we
investigated shapes other than rectangular, namely a contracting and
an expanding one, and compared their cruising velocities. We found
that the contracting shape is the best, and postulate that this may
possibly be explained by the tip vortices quickly ``loosing touch''
to the swimmer, which reduces their influence on the drag. 
\textcolor{black}{Caudal
fins resembling the expanding shape are found in many fish (thunniform
fin), while the contracting form is found in some amphibians (protoceral
form).}

\paragraph*{\textcolor{black}{\footnotesize{}Acknowledgments}}

\noindent \textcolor{black}{\footnotesize{}Financial support from the
ANR (Grant 15-CE40-0019) and DFG (Grant SE 824\textbackslash{}26 -1)
is gratefully acknowledged and CPU time from IDRIS, project i20152a1664.
For this work we were also granted access to the HPC resources of
Aix-Marseille Université financed by the project Equip@Meso (ANR-10-EQPX-29-01).
TE, KS, MF, FL and JS thankfully acknowledge financial support granted
within the French-German PROCOPE project FIFIT. DK gratefully acknowledges
the financial support from the JSPS (KAKENHI Grant Number 
15F15061).}{\footnotesize \par}

\appendix
%dummy comment inserted by tex2lyx to ensure that this paragraph is not empty

\section{Validation: Thrust generated by a heaving plate}

As validation case we consider the numerical work presented in \cite{Yeh2014}.
They consider a flexible panel with an imposed heaving motion at the
leading edge with zero angle of attack (i.e., no pitching imposed).
The authors study the influence of the driving frequency, normalized
by the eigenfrequency in fluid, $\phi=f/f_{1,f}$ by varying the elastic
properties of the flexible plate. Two regimes are identified, which
maximize either the velocity for $\phi\approx1.1$ or the efficiency
for $\phi\approx1.7$. The reference solution is computed using the
Lattice-Boltzmann method, which approximates incompressible fluid
flow without solving an elliptical Poisson equation. An overset approach
with a refined region in the vicinity of the plate and a coarser one
in the far field are used. The flexible plate is modeled using the
lattice-spring model, i.e., it is approximated as a system of mass
points connected by springs.

Here, we consider only a single simulation out of the dataset presented
in \cite{Yeh2014}, arbitrarily fixing the added mass parameter (as
defined in \cite{Yeh2014}) to $T=1$ and the frequency ratio to $\phi=1.1$.
The non-dimensional solid properties in our model are $\mu=0.4031$
and $\eta=1.6669$. The plate's width is $w=0.4$, and the prescribed
heaving motion is given as $y=a_{0}\cos\left(2\pi ft\right),$ where
$a_{0}=0.1$. The Reynolds number is $Re=2\pi f\ell a_{0}/\nu=250$.
The domain size is $4\times4\times2$ and it is discretized using
$768\times768\times384$ points, with the penalization parameter equal
to $C_{\eta}=2\cdot10^{-4}$. We apply a vorticity sponge to remove
the periodicity. The validation is performed in the steady cruising
state. The handling of the mean flow is the same as above, using 
$m_{\mathrm{fluid}}=0.5$.
In coarser resolution pre-runs, the mean flow was started from rest
and the terminal value reached in that simulation, $1.3$, was set
as initial condition $u_{\infty}\left(t=0\right)$ in the high-resolution
case, to further reduce the computational cost. In the high-resolution
case, $11$ cycles were performed.

Table (\ref{tab:Cycle-averaged-results-for}) compares the results
obtained during the last cycle with those given in \cite{Yeh2014}.
The cruising speed is slightly reduced in the present work, but the
agreement to within 5\% is still tolerable. The aerodynamic power,
\textcolor{black}{which can be computed as 
$P_{\mathrm{aero}}=\int\underline{u}_{s}\left(\underline{u}-\underline{u}_{s}
\right)/C_{\eta}$
when using the penalization method \cite{Engels2015a}}, is overpredicted
by about 25\%. Both, overprediction of power requirement and underprediction
of cruising speed are related to an elevated drag coefficient owing
to the smoothing layer in the $\chi$-function. The trailing edge
displacement $d$ is very close to the reference solution (1\%).

\begin{figure}
\noindent \begin{centering}
\includegraphics[width=1\columnwidth]{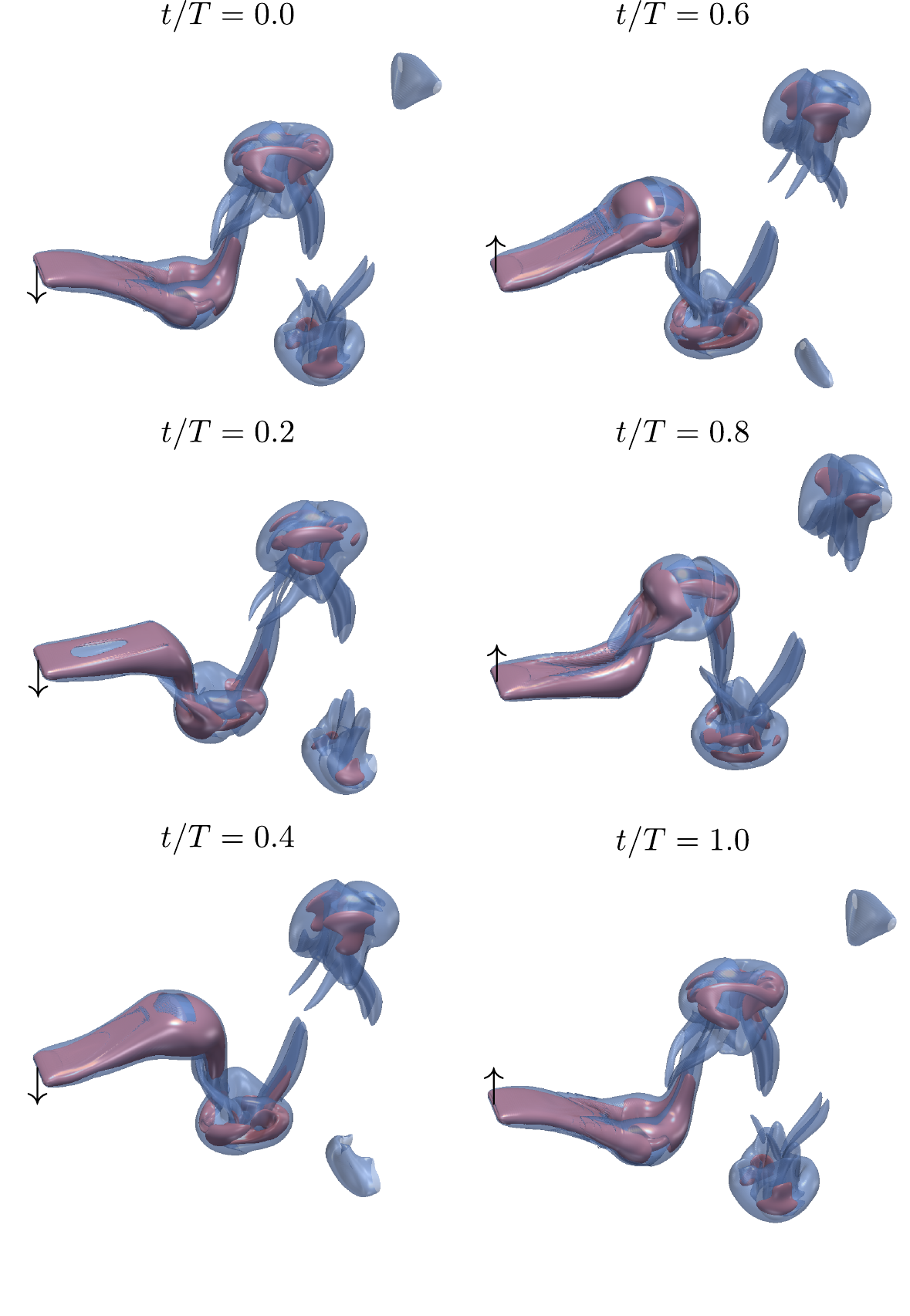} 
\par\end{centering}
\caption{\emph{\label{fig:heaving_flow}Flow generated by the heaving plate,
visualized by isosurfaces of vorticity magnitude. Blue semi-transparent
corresponds to $\left\Vert \underline{\omega}\right\Vert =5$, purple
to $\left\Vert \underline{\omega}\right\Vert =20$. The Reynolds number
is 250.}}
\end{figure}

The flow field is visualized in figure \ref{fig:heaving_flow}. At
each half-stroke, a vortex is shed, which travels perpendicular to
the mean flow on a V-shaped path. The flow field is qualitatively
similar to the reference computation.

In conclusion, the comparison yields reasonable agreement with the
literature, and the remaining difference can be attributed to differences
in modeling. The present plate is rigid in the lateral direction,
while the reference data is obtained with a 2D flexible plate. This
simplification has also an influence on the solid model parameters
$\eta$ and $\mu$. The former is the non-dimensional rigidity, corrected
by the Poisson ratio of the solid material \cite{Landau1986}. Owing
to this correction, the eigenfrequency in vacuum of the beam, $f_{0}=3.516/2\pi\,\,\sqrt{\eta/\mu}$
is slightly different (6\%) from the corresponding eigenfrequency
of the 2D flexible plate. However, the differences between both studies
are reasonable and we thus conclude that our fluid\textendash structure
interaction module is validated. 
\begin{table}
\noindent \begin{centering}
\begin{tabular*}{1\columnwidth}{@{\extracolsep{\fill}}@{\extracolsep{\fill}}ccccc}
\toprule 
 & $u_{\infty}$  & $P_{\mathrm{aero}}$  & $\eta_{\mathrm{eff}}$  & $d$\tabularnewline
\midrule
\midrule 
Yeh \& Alexeev \cite{Yeh2014}  & 1.47  & 5.11  & 0.29  & 7.60\tabularnewline
\midrule 
Present  & 1.40  & 6.53  & 0.22  & 7.68\tabularnewline
\bottomrule
\end{tabular*}
\par\end{centering}
\caption{\emph{Cycle-averaged results for the heaving plate.\label{tab:Cycle-averaged-results-for}}}
\end{table}
 
\section*{\textemdash \textemdash \textemdash \textemdash \textemdash \textendash{}}

\bibliographystyle{elsarticle-harv}
\phantomsection\addcontentsline{toc}{section}{\refname}\bibliography{bibliography}

\begin{thebibliography}{20}
\expandafter\ifx\csname natexlab\endcsname\relax\def\natexlab#1{#1}\fi
\expandafter\ifx\csname url\endcsname\relax
  \def\url#1{\texttt{#1}}\fi
\expandafter\ifx\csname urlprefix\endcsname\relax\def\urlprefix{URL }\fi

\bibitem[{Alben(2009)}]{Alben2009}
Alben, S., 2009. Simulating the dynamics of flexible bodies and vortex sheets.
  J. Comput. Phys. 228, 2587--2603.

\bibitem[{Dewey et~al.(2013)Dewey, Boschitsch, Moored, Stone, and
  Smits}]{Dewey2013}
Dewey, P.~A., Boschitsch, B.~M., Moored, K.~W., Stone, H.~A., Smits, A.~J.,
  2013. Scaling laws for the thrust production of flexible pitching panels. J.
  Fluid Mech. 732, 29--46.

\bibitem[{Ehrenstein et~al.(2014)Ehrenstein, Marquillie, and
  Eloy}]{Ehrenstein2014}
Ehrenstein, U., Marquillie, M., Eloy, C., 2014. Skin friction on a flapping
  plate in uniform flow. Phil. Trans. R. Soc. A 372.

\bibitem[{Eloy(2012)}]{Eloy2012}
Eloy, C., 2012. Optimal strouhal number for swimming animals. J. Fluids Struct.
  30, 205--218.

\bibitem[{Engels et~al.(2012)Engels, Kolomenskiy, Schneider, and
  Sesterhenn}]{Engels2012a}
Engels, T., Kolomenskiy, D., Schneider, K., Sesterhenn, J., 2012.
  Two-dimensional simulation of the fluttering instability using a
  pseudospectral method with volume penalization. Computers \& Structures 122,
  101--112.

\bibitem[{Engels et~al.(2015)Engels, Kolomenskiy, Schneider, and
  Sesterhenn}]{Engels2014}
Engels, T., Kolomenskiy, D., Schneider, K., Sesterhenn, J., 2015. Numerical
  simulation of fluid-structure interaction with the volume penalization
  method. J. Comput. Phys. 281, 96--115.

\bibitem[{Engels et~al.(2016)Engels, Kolomenskiy, Schneider, and
  Sesterhenn}]{Engels2015a}
Engels, T., Kolomenskiy, D., Schneider, K., Sesterhenn, J., 2016. {F}lu{SI}:
  {A} novel parallel simulation tool for flapping insect flight using a
  {F}ourier method with volume penalization. SIAM J. Sci. Comput. 38~(5),
  S3--S24.

\bibitem[{Eysden and Sader(2007)}]{Sader2007}
Eysden, C. A.~V., Sader, J.~E., 2007. Frequency response of cantilever beams
  immersed in viscous fluids with applications to the atomic force microscope:
  {A}rbitrary mode order. J. Appl. Phys. 101, 044908.

\bibitem[{Kang et~al.(2011)Kang, Aono, Cesnik, and Shyy}]{Kang2011}
Kang, C.-K., Aono, H., Cesnik, C. E.~S., Shyy, W., 2011. Effects of flexibility
  on the aerodynamic performance of flapping wings. J. Fluid Mech. 689, 32--74.

\bibitem[{Kolomenskiy et~al.(2013)Kolomenskiy, Engels, and
  Schneider}]{Kolomenskiy2013}
Kolomenskiy, D., Engels, T., Schneider, K., 2013. Numerical modelling of
  flexible heaving foils. J. Aero Aqua Bio-mechanisms 3, 22--28.

\bibitem[{Landau and Lifshitz(1986)}]{Landau1986}
Landau, L., Lifshitz, E., 1986. Theory of Elasticity, 3rd Edition. Vol.~7 of
  Theoretical Physics. Butterworth-Heinemann.

\bibitem[{Lighthill(1971)}]{Lighthill1971}
Lighthill, M.~J., 1971. Large-amplitude elongated-body theory of fish
  locomotion. Proc. R. Soc. Lond. B 179, 125--138.

\bibitem[{Michelin et~al.(2008)Michelin, Smith, and Glover}]{Michelin2008}
Michelin, S., Smith, S.~L., Glover, B., 2008. Vortex shedding model of a
  flapping flag. J. Fluid Mech. 617, 1--10.

\bibitem[{Ramananarivo et~al.(2011)Ramananarivo, {Godoy-Diana}, and
  Thiria}]{Ramananarivo2011}
Ramananarivo, S., {Godoy-Diana}, R., Thiria, B., 2011. Rather than resonance,
  flapping wing flyers may play on aerodynamics to improve performance. Proc.
  Natl. Acad. Sci. USA 108, 5964--5969.

\bibitem[{Raspa et~al.(2014)Raspa, Ramananarivo, Thiria, and
  Godoy-Diana}]{Raspa2014}
Raspa, V., Ramananarivo, S., Thiria, B., Godoy-Diana, R., 2014. Vortex-induced
  drag and the role of aspect ratio in undulatory swimmers. Phys. Fluids
  26~(4).

\bibitem[{Schneider(2005)}]{Schneider2005}
Schneider, K., 2005. Numerical simulation of the transient flow behaviour in
  chemical reactors using a penalisation method. Computers \& Fluids 34,
  1223--1238.

\bibitem[{Triantafyllou et~al.(1993)Triantafyllou, Triantafyllou, and
  Grosenbaugh}]{Triantafyllou1993}
Triantafyllou, G.~S., Triantafyllou, M.~S., Grosenbaugh, M.~A., 1993. Optimal
  thrust development in oscillating foils with application to fish propulsion.
  J. Fluid Struct. 7, 205--224.

\bibitem[{Triantafyllou et~al.(2004)Triantafyllou, Techet, and
  F.S.Hover}]{Triantafyllou2004}
Triantafyllou, M., Techet, A., F.S.Hover, 2004. Review of experimental work in
  biomimetic foils. IEEE J. Oceanic Eng. 29, 585--594.

\bibitem[{Vanella et~al.(2009)Vanella, Fitzgerald, Preidikman, Balaras, and
  Balachandran}]{Vanella2009}
Vanella, M., Fitzgerald, T., Preidikman, S., Balaras, E., Balachandran, B.,
  2009. Influence of flexibility on the aerodynamic performance of a hovering
  wing. J. Exp. Biol 121, 95--105.

\bibitem[{Yeh and Alexeev(2014)}]{Yeh2014}
Yeh, P.~D., Alexeev, A., 2014. Free swimming of an elastic plate plunging at
  low {R}eynolds number. Phys. Fluids 26~(5), --.

\end{thebibliography}

\end{document}